\documentclass[aps,prev,twocolumn,preprintnumbers,floatfix,nofootinbib]{revtex4-1}
\usepackage{graphicx}
\usepackage{bm}
\usepackage{times}
\usepackage{slashed}
\usepackage{color}
\usepackage{color}
\def\br{\begin{eqnarray}}
\def\er{\end{eqnarray}}
\def\be{\begin{equation}}
\def\ee{\end{equation}}

\begin{document}

\title{Excluding the Light Dark Matter Window of a 331 Model Using LHC and Direct Dark Matter Detection Data} 

\author{D. Cogollo$^{a}$}\email{diegocogollo@df.ufcg.edu.br}
\author{Alma X. Gonzalez-Morales$^{b}$}\email{alxogonz@ucsc.edu}
\author{Farinaldo S. Queiroz$^{b}$}\email{fdasilva@ucsc.edu}
\author{P. Rebello Teles$^{c}$}\email{patricia.rebello.teles@cern.ch}

\affiliation{
$^{a}$Departamento de Física, Universidade Federal de Campina Grande,
Caixa Postal 10071, 58109-970, Campina Grande, PB, Brazil\\
$^{b}$Department of Physics and Santa Cruz Institute for Particle Physics
University of California, Santa Cruz, CA 95064, USA\\
$^{c}$CERN, PH Department, 1211 Geneva, Switzerland}

\pacs{}
\date{\today}
\vspace{1cm}

\begin{abstract}

We sift the impact of the recent Higgs precise measurements, and recent dark matter direct detection results, on the dark sector of an electroweak extension of the Standard Model that has a complex scalar as dark matter. We find that in this model the Higgs decays with a large branching ratio into dark matter particles, and charged scalars when these are kinematically available, for any coupling strength differently from the so called Higgs portal. Moreover, we compute the abundance and spin-independent WIMP-nucleon scattering cross section, which are driven by the Higgs and $Z^{\prime}$ boson processes. We decisively exclude the $1-500$~GeV dark matter window and find the most stringent lower bound in the literature on the scale of symmetry breaking of the model namely $10$~TeV, after applying the LUX-2013 limit. Interestingly, the projected XENON1T constraint will be able to rule out the entire $1$~GeV-$1000$~GeV dark matter mass range. Lastly, for completeness, we compute the charged scalar production cross section at the LHC and comment on the possibility of detection at current and future LHC runnings. 
\end{abstract}
    
\maketitle

\section{INTRODUCTION}
\label{introduction}

The nature of the dark matter (DM) is one of the greatest puzzles in current science, once the DM constitutes approximately $23\%$ of the Universe budget. There are promising ongoing searches aimed to detect and find the nature of the DM that permeates the Universe. There are many dark matter candidates in the literature, but the most seemingly promising ones are the so called WIMPs (Weakly Interacting Massive Particles) for having a thermal cross section at the electroweak scale, naturally addressing the structure formation process, and being predicted in many interesting particle physics models.

There are four different methods to infer the presence or detect theses WIMPs known as indirect detection, direct detection, colliders and cosmological observations. Indirect detection searches have found some excess events in the gamma-ray emission \cite{gammaray} and in the cosmic ray emission \cite{cosmicray} which might be explained by annihilation of WIMPs in our galaxy \cite{IDexplanations}, which is in contradiction with recent dwarf galaxies constraints \cite{dwarfsAlma}. Likewise, some direct detection experiments such CoGeNT \cite{cogent}, DAMA \cite{dama}, CRESST \cite{cresst} and most recently CDMSII-Si \cite{cdms} have observed some excess events consistent with WIMP scatterings \cite{ddwimps}. Due to some possible leakage of background events into the signal region at low energies and the non-observation of such events in the XENON \cite{xenon} and LUX \cite{lux} experiments, those events do not constitute an irrefutable DM signal \cite{backgrounds}. Furthermore, there are cosmological measurements of the Cosmic Microwave Background that revealed some degree of dark radiation observed in the Planck data \cite{planck}, among other satellites, \cite{satelites} that may constitute an evidence for a sub-dominant non-thermal production of DM \cite{nonthermal}. Lastly, collider data, which provide an important and complementary method to infer the nature of the dark matter have not observed any positive signal for a stable particle and just bounds on the mass and coupling strengths had been derived \cite{Willshepherd}.

In this work we will focus on a compelling extension of the Standard Model (SM) namely 331LHN, that might address these evidences. 331LHN stands for a electroweak extension of the SM where doublets are replaced by triplets, both in the scalar and the fermion sector. This proposal has been able to endure all electroweak precise measurements and reproduce the SM results concerning the Higgs signal strength \cite{331DM2} as oppose to other 331 model extensions which predict a $H\rightarrow \gamma \gamma$ enhancement \cite{hgamma}. It also has a rich particle spectrum comprised of charged scalars , gauge bosons, sterile neutrinos and exotic quarks, with interesting phenomenological aspects, which had been investigated elsewhere \cite{gaugebosonsLHC}. Furthermore, this model does have a plausible DM candidate able to explain the gamma-ray excess observed in the Fermi-LAT data at the Galactic Center \cite{gammaray} differently to other versions \cite{331minimal,331econo,331susyecono,331two,331susyecono2,331g2muon,Caetano:2012qc,Ferreira:2013nla,Dias:2013kma,Dong:2014wsa}; and offers a plausible mechanism to account for the dark radiation observed by the Planck Collaboration, through a sub-dominant non-thermal production of WIMPs,  while evading structure formation, Big Bang Nucleosynthesis and CMB bounds among others \cite{331darkradiation}. An extensive analysis concerning the heavy fermions present in the model has been done, and stringent bounds on the mass of the lightest sterile neutrino have been found as a function of the $Z^{\prime}$ mass in Ref.\cite{331DM3}, and in a model independent fashion in Ref.\cite{DarkZp}. It is important to stress that such constraints on the $Z^{\prime}$ mass {\it do apply, at some level, to all 331 models}, that have fermions as DM candidates, as discussed in Fig.7 of Ref.\cite{331DM3}, and are complementary to others coming from colliders \cite{331collider,331collider2}, FCNC \cite{331FCNC}, muon decay \cite{331muon}, top decay \cite{331top} analyses, and oblique STU parameters \cite{Liu:1993fwa,Martinez:2009ik}.

That being said, here we will discuss the 331LHN model 
 which has two not co-existing DM candidates \cite{331DM2,331DM1}. Our purpose is to derive constraints on the dark sector of this well motivated model in the light of the present bounds in the Higgs signal strength and DM observables. In particular, we will exclude DM masses below $500$~GeV, and discuss the impact of this exclusion on the dark sector of the model with focus on the charged scalar which predicted in this model.

The paper is organized as follows: in section \ref{sec1} we briefly introduce the 331LHN model. In section \ref{darksector} we derive bounds on the dark sector of the model, and in section \ref{LHC} we comment about the possibility of detection at current and future LHC runnings. Finally we present our conclusions in section \ref{sec:conclusion}

\section{The 3-3-1LHN Model}
\label{sec1}

As we mentioned before, 3-3-1 stands for an extension of the electroweak sector of the SM where the electroweak sector $SU(2)_L \otimes U(1)_Y$ is enlarged to $SU(3)_L \otimes U(1)_N$. As a result the doublets in the electroweak sector of the SM are replaced by triplets. This extension is motivated by important matters not fully addressed by the SM , namely  the number of generations, the neutrinos masses, and the lack of a plausible DM candidate. Moreover, it reproduces precisely the SM results, including the Higgs properties as shown in Ref.\cite{331DM2}. Hence, the 3-3-1LHN remains as a compelling extension of the SM. In what follows, we will not dwell on unnecessary details but shortly review the key points of this model, which will allow the reader to follow our reasoning.

\subsection*{Leptonic Sector}

The leptons are displayed in triplet and singlet representations as follows:
\begin{eqnarray}
f_{aL} & = & \left (
\begin{array}{c}
\nu_{a} \\
e_{a} \\
N_{a}
\end{array}
\right )_L\sim(1\,,\,3\,,\,-1/3)\nonumber\\
     &e_{aR}& \sim(1,1,-1)\,,\, N_{aR}\,\sim(1,1,0),
\label{L}
\end{eqnarray}
where $a=1,2,3$ runs over the three lepton families, and $N_{a(L,R)}$ are the heavy fermions added to the SM particle spectrum. The shortened representation $(1\,,\,3\,,\,-1/3)$ simply refers to the quantum numbers of the symmetry group $SU(3)_c \otimes SU(3)_L\otimes U(1)_N$. 

The SM mass spectrum will be reproduced. In particular, the charged leptons will acquire mass terms through the first term of the Yukawa Lagrangian in Eq.(\ref{yukawa1}), whereas the neutrinos through a dimension 5 effective operators according to Eq.(\ref{numasses}).

\begin{equation}
{\cal L}^Y \supset G_{ab}\bar f_{aL} \rho e_{bR}+g^{\prime}_{ab}\bar{f}_{aL}\chi N_{bR}+ \mbox{h.c}, 
\label{yukawa1}
\end{equation}

\begin{equation}
{\cal L}^Y \supset \frac{y_{ab}}{\Lambda}\bar{f^c}_{aL}\eta^{\star}\eta^{\dagger} f_{bL} + \mbox{h.c},
\label{numasses}
\end{equation}where $\rho, \eta$ and $\chi$ are the scalar triplets introduced in Eq.(\ref{scalarcont}). 

We do not show explicitly the masses of the SM particles in this work and just present the mass of the heavy fermions ($N_a$) introduced by the 3-3-1 symmetry as follows, 

\be
M_{N_a}=\frac{g^{\prime}_{aa}}{\sqrt{2}}v_{\chi^{\prime}}\,,
\label{mneut}
\ee
where $g^{\prime}_{aa}$ are the Yukawa couplings that appear in the last term of Eq.(\ref{yukawa1}). We assume all Yukawa couplings to be diagonal with a normal hierarchy throughout this work. The hierarchy adopted does not lead to any impact on our conclusions.

\subsection*{Hadronic Sector}

The quarks in the theory are also arranged in triplets. The third generation lives in a triplet representation while the other two generations are in anti-triplet representations of $SU(3)_L$, so that triangle anomalies are canceled as follows~\cite{331minimal},
\begin{eqnarray}
&&Q_{iL} = \left (
\begin{array}{c}
d_{i} \\
-u_{i} \\
q^{\prime}_{i}
\end{array}
\right )_L\sim(3\,,\,\bar{3}\,,\,0)\,, \nonumber \\
&&
u_{iR}\,\sim(3,1,2/3),\,\,\,
\,\,d_{iR}\,\sim(3,1,-1/3)\,,\,\,\,\, q^{\prime}_{iR}\,\sim(3,1,-1/3),\nonumber \\
&&Q_{3L} = \left (
\begin{array}{c}
u_{3} \\
d_{3} \\
q^{\prime}_{3}
\end{array}
\right )_L\sim(3\,,\,3\,,\,1/3)\,, \nonumber \\
&&
u_{3R}\,\sim(3,1,2/3),
\,\,d_{3R}\,\sim(3,1,-1/3)\,,\,q^{\prime}_{3R}\,\sim(3,1,2/3)
\label{quarks} 
\end{eqnarray}
where the index $i=1,2$ refers to the first two generations. The primed quarks $(q^{\prime})$ are heavy quarks with the following electric charges, $Q(q^{\prime}_1)= -1/3 ,Q(q^{\prime}_2)=-1/3, Q(q^{\prime}_3)=2/3$. These quarks do not couple with the SM gauge bosons but couple with the extra gauge bosons introduced by the 3-3-1 symmetry that we will discuss further \footnote{As for quark physics studies we refer to Refs.\cite{quarkstudies}.}

The masses of all quarks are derived from the Yukawa Lagrangian in Eq.(\ref{yukawa2}),

\begin{eqnarray}
&-&{\cal L}^Y  \supset \alpha_{ij} \bar Q_{iL}\chi^* q^{\prime}_{jR} +f_{33} \bar Q_{3L}\chi q^{\prime}_{3R} + g_{ia}\bar Q_{iL}\eta^* d_{aR} \nonumber \\
&&+h_{3a} \bar Q_{3L}\eta u_{aR} +g_{3a}\bar Q_{3L}\rho d_{aR}+h_{ia}\bar Q_{iL}\rho^* u_{aR} + \mbox{h.c}., \nonumber \\
\label{yukawa2}
\end{eqnarray}with $i,j=1,2$. and $a=1,2,3$.

Again, the SM quarks masses are equal to the usual ones, once $v_{\rho}=v_{\eta}=v$, where $v=v_{SM}/\sqrt{2}$~GeV \footnote{One might consider scenarios where $v_{\rho} \neq v_{\eta}$, and in those setups different conclusions might be found.}. As for the three new quarks $q^{\prime}_a$ they have their masses given by the first two terms of Eq.(\ref{yukawa2}) with,

\be
M_{q^{\prime}_a}=\frac{\alpha_{aa}}{\sqrt{2}}v_{\chi^{\prime}}\,.
\label{mquarks}
\ee

One can clearly see that the masses of the new quarks are proportional to the scale of symmetry breaking of the model, which we assumed to lie at the TeV scale. Anyway, the new quarks do not play any role in the current work and will be thus completely ignored henceforth.

\subsection*{Gauge Bosons}
\label{gaugebosons}

Due to the enlarged electroweak gauge group, $SU(2)_L \rightarrow SU(3)_L$, extra gauge bosons will arise in the 3-3-1LHN model, namely: $Z^{\prime}, W^{\prime \pm},$ and $U^{0}$ and $U^{0\dagger}$. These bosons have masses proportional to the scale of symmetry breaking of the model as follows,
\begin{eqnarray}
M^2_{Z^\prime} & = & \frac{g^{2}}{4(3-4s_W^2)}[4c^{2}_{W}v_{\chi^\prime}^2 +\frac{v^{2}}{c^{2}_{W}}+\frac{v^{2}(1-2s^{2}_{W})^2}{c^{2}_{W}}]
\nonumber \\
M^2_{W^{\prime}} & = & M^2_{U^0} = \frac{1}{4}g^2(v_{\chi^\prime}^2+v^2)\,,
\label{massvec}
\end{eqnarray}where we used the shortened notation $\sin_{\theta_W} =s_W$ and  $\cos_{\theta_W} =c_W$. Notice that their masses are also balanced by the scale of symmetry breaking of the model ($v_{\chi}^{\prime}$).

These gauge bosons give rise to the neutral and charged current below,

\begin{eqnarray}
\label{NCeq}
&&{\cal L}^{NC} =-\frac{g}{2 \cos\theta_W}\sum_{f} \Bigl[\bar
f\,  \gamma^\mu\ (g^\prime_V + g^\prime_A \gamma^5)f \, { Z_\mu^\prime}\Bigr],
\end{eqnarray}

\br
{\cal L}_{NH} &=&
-\frac{g}{\sqrt{2}}\left[\bar{\nu}^a_L\gamma^\mu e_L^a W^+_\mu +\bar{N}_L^a\gamma^\mu e_L^a W^{\prime +}_{\mu} + \bar{\nu}^a_L\gamma^\mu N_L^a U^0_\mu  \right. \nonumber \\
&& \left. +\left(\bar{u}_{3L}\gamma^\mu d_{3L}  +\bar{u}_{iL}\gamma^\mu d_{iL}\right)W^+_{\mu}  \right. \nonumber \\
&& \left. +\left(\bar{q}^\prime_{3L}\gamma^\mu d_{3L}  +\bar{u}_{iL}\gamma^\mu q^\prime_{iL}\right)W^{\prime +}_{\mu} \right. \nonumber \\
&& \left. +\left(\bar{u}_{3L}\gamma^\mu q^\prime_{3L}  -\bar{q}^\prime_{iL}\gamma^\mu d_{iL}\right)U_\mu^0
+ {\mbox h.c.}
\right]\,
\label{CC} 
\er  where ($g^{\prime}_V$) and ($g^{\prime}_A$) are the vector and axial couplings with quarks/leptons as shown in \cite{331DM3}. Now we presented the masses and the current involving these gauge bosons we discuss the current collider and electroweak constraints.

\subsection*{LHC and Electroweak Constraints}

Stringent bounds on the mass of these bosons can be found in the literature. We will rigorously adopt them throughout this work \cite{gaugebosonsLHC}. However we would like to mention that the $Z^{\prime}$ {\it does not} couple to the SM fermions in the same way the Z boson does. In fact, the couplings of the $Z^{\prime}$ with the SM quarks and charged leptons are dwindled in $\sim 50\%$, while with SM neutrinos are $80\%$ suppressed in comparison with the respective SM Z couplings ones. In other words, the general neutral current written in Eq.(\ref{NCeq}) has vector and axial couplings with quarks, and leptons, suppressed in comparison with the Z couplings aforesaid.

It is important to emphasize this fact because recent solid limits were derived on the mass of the $Z^{\prime}$ boson for the 3-3-1 model with right handed neutrinos using CMS data: $M_{Z^{\prime}} > 2.2$~TeV ~\cite{331collider}. However, this constraint does not directly apply to our model because the $Z^{\prime}$ decays mostly into missing energy (heavy neutral fermions). For the regime where $M_{N_{a}}< M_{Z^{\prime}}/2$, the $Z^{\prime}$ decays at $100\%$ into fermion pairs ($\overline{N_a}N_a$) as opposed to Ref.\cite{331collider}, which assumed that the $Z^{\prime}$ decays primarily into quarks and charged leptons. Nevertheless, when $\overline{N_a}N_a$ channel is not kinematically accessible, 
the results found in Ref.~\cite{331collider} do apply to our model. Either way, as we mentioned earlier we will always take this face value limit throughout this work. For complete analyses concerning the phenomenology of this neutral boson see Ref.\cite{gaugebosonsLHC}.

As for the gauge bosons present in the charged current, there is a lack of collider bounds on the mass of the gauge boson $U^0$. Albeit, since the mass terms of $W^{\prime}$ and $U^0$ bosons are the same, according to Eq.(\ref{massvec}), any constrain found on the mass of the  $W^{\prime}$ is applicable to $U^0$ as well. The $W^{\prime}$ has been vastly searched at the LHC \cite{Wprime1,Wprime2}: from LEP-II we have $M_{W^{\prime}} > 105$~GeV, because this charged boson could have been easily produced via drell Yan processes; and from the ATLAS Collaboration we know that a $W^{\prime}$ boson has been ruled out for $M_{W^{\prime}} < 2.55$~TeV at $95\%$ C.L, assuming SM coupling with fermions. Similarly to the $Z^{\prime}$ case, we {\it will strictly use the face value bound from ATLAS}, but we would like to stress that this limit does not directly apply to our model for the following reasons: 

(i) The boson $W^{\prime}$ does not couple similarly to the SM W boson as can be seen in Eq.(\ref{CC}). 

(ii) $W^{\prime}$ decays predominantly into sterile neutrino plus electron ($N e$) pairs; 

(iii) In proton-proton collisions, the $W^{\prime}$ production is different from the W one. There are other processes in addition to Drell-Yan processes that contribute, such as a t-channel process mediated by new quark $q^{\prime}_1$, and three s-channel processes mediated by the Higgs, the scalar $S_2$ and the $Z^{\prime}$. 

Therefore one cannot straightforwardly apply the $Z^{\prime}$ and $W^{\prime}$ limits into this model. Anyhow, at which degree these bounds are applicable to the 331LHN is far beyond the scope of this paper 
but we will be conservative and adopt those limits in the present analysis.

 In summary the LHC bounds read:

\begin{itemize}
\item $M_{Z^{\prime}} > 2.2$~TeV,
\item $M_{W^{\prime}} > 2.55$~TeV.
\end{itemize}

Those limits can be translated into $v_{\chi^{\prime}} > 5.5$~TeV, which will be  respected throughout since in the forthcoming results we use $v_{\chi^{\prime}} \geq 8$~TeV.
Additional limits coming from electroweak precision such as those from STU oblique parameters do not offer competitive bounds \cite{Liu:1993fwa,Martinez:2009ik}

Here we aim to derive lower limits on the mass of the charged scalars of the model, which could be lighter than the mass of this boson at the cost of some tuning in the couplings, as we shall see in the next section. 

\subsection*{Scalar Content}
\label{scalarcontent}

The symmetry breaking pattern $ SU(3)_L\otimes U(1)_N \rightarrow SU(2)_L\otimes U(1)_Y$ $\rightarrow$ $U(1)_{QED}$ is accomplished by three scalar triplets, namely
 
\begin{eqnarray}
\eta = \left (
\begin{array}{c}
\eta^0 \\
\eta^- \\
\eta^{\prime 0}
\end{array}
\right ),\,\rho = \left (
\begin{array}{c}
\rho^+ \\
\rho^0 \\
\rho^{\prime +}
\end{array}
\right ),\,
\chi = \left (
\begin{array}{c}
\chi^0 \\
\chi^{-} \\
\chi^{\prime 0}
\end{array}
\right )\,,
\label{scalarcont} 
\end{eqnarray}
which form the following scalar potential,

\begin{eqnarray} V(\eta,\rho,\chi)&=&\mu_\chi^2 \chi^2 +\mu_\eta^2\eta^2
+\mu_\rho^2\rho^2+\lambda_1\chi^4 +\lambda_2\eta^4
+\lambda_3\rho^4+ \nonumber \\
&&\lambda_4(\chi^{\dagger}\chi)(\eta^{\dagger}\eta)
+\lambda_5(\chi^{\dagger}\chi)(\rho^{\dagger}\rho)+\lambda_6
(\eta^{\dagger}\eta)(\rho^{\dagger}\rho)+ \nonumber \\
&&\lambda_7(\chi^{\dagger}\eta)(\eta^{\dagger}\chi)
+\lambda_8(\chi^{\dagger}\rho)(\rho^{\dagger}\chi)+\lambda_9
(\eta^{\dagger}\rho)(\rho^{\dagger}\eta) \nonumber \\
&&-\frac{f}{\sqrt{2}}\epsilon^{ijk}\eta_i \rho_j \chi_k +\mbox{H.c}.
\label{potential}
\end{eqnarray}
with $\eta$ and $\chi$ both transforming as $(1\,,\,3\,,\,-1/3)$ while $\rho$ as $(1\,,\,3\,,\,2/3)$ under $SU(3)_c \otimes SU(3)_L \otimes U(1)_N$ and $f$ assumed to be equal to $v_{\chi^ {\prime}}$.

The scalar triplets above are invoked in order to generate masses for all fermions in the model after the spontaneous symmetry breaking mechanism represented by the non-zero vacuum expectation value (vev), of the scalars $\eta^0, \rho^0\ \mbox{and}\ \chi^{\prime 0}$ as,
\begin{eqnarray}
 \eta^0 , \rho^0 , \chi^{\prime 0} \rightarrow  \frac{1}{\sqrt{2}} (v_{\eta ,\rho ,\chi^{\prime}} 
+R_{ \eta ,\rho ,\chi^{\prime}} +iI_{\eta ,\rho ,\chi^{\prime}})\,.
\label{vacua} 
\end{eqnarray}
There are additional neutral scalars in the spectrum, namely $\eta^{\prime 0}$ and $\chi^{0}$, which are enforced not to develop  vev's  in order to preserve the discrete symmetry given by,
\begin{eqnarray}
(N_L\,,\,N_R\,,\,d^{\prime}_i\,,\,u^{\prime}_3\,,\,\rho^{\prime +}\,,\,\eta^{\prime 0}\,,\,\chi^{0}\,,\,\chi^-\,,\, V^+\,,\,U^{0 \dagger}) \rightarrow -1.\nonumber\\
\label{discretesymmetryI}
\end{eqnarray}
where $d^{\prime}_i$ and $u^{\prime}_3$ are new heavy quarks predicted in the model due to the enlarged gauge group. The remaining fields all transforms trivially under this symmetry.  We indicate it with $P=(-1)^{3(B-L)+2s}$, where $B$ is the baryon number, $L$ is the lepton number and $s$ is spin of the field; this parity symmetry can be understood as a R-parity symmetry like the one in the minimal supersymmetric standard model. Note that the heavy fermions (N's) do not carry a lepton number.

This discrete symmetry induce three distinct consequences. First, it stabilizes the lightest particle charged under the symmetry. Second, it simplifies the scalar mass spectrum of the model. Last but not least, it prohibits Yukawa mass terms that would mix the new quarks with the SM ones. The downside is that we rely on the assumption that the remaining neutral scalars $\eta^{\prime 0}$ and $\chi^{0}$ do not develop a vev. This is a crucial assumption in what follows, and an important discussion on this topic has been given in Refs.\cite{331DM3,3311,DMeconomic2}. Moreover, a more elegant way to explain the WIMP stability would be gauging this discrete symmetry as discussed in Ref.\cite{3311}. Less appealing DM scenarios in 331 models have been studied elsewhere \cite{DMelsewhere}. 

In the 3-3-1LHN model there are two possible DM candidates: a complex scalar $\phi$ (the mass eigenstate resulting from $\eta^{0 \prime}$ and $\chi^0$), and a heavy fermion $N_i$ (the lightest of the new heavy fermions). We will restrain ourselves to the case where the scalar is the lightest particle,  protected by the parity symmetry. We investigate its consequences on the dark sector of the model under the assumption that such scalar is a plausible DM candidate, i.e. it must be able to reproduce the DM abundance, as well as satisfy the direct detection bounds.
Anyhow, once the pattern of symmetry breaking has been established one can straightforwardly derive the mass eigenstates of the model. After  spontaneous symmetry breaking the {\it three CP-even neutral scalars} mass eigenstates ($H,S_1,S_2$) are found to be,
\begin{eqnarray}
M^{2}_{S_{1}} & = & \frac{v^{2}}{4}+2v_{\chi^\prime}^{2}\lambda_{1}\,, \nonumber \\
M^{2}_{S_{2}} & = & \frac{1}{2}(v_{\chi^\prime}^{2}+2v^{2}(\lambda_{2}+\lambda_{3}+\lambda_{6}))\,, \nonumber \\
M^{2}_{H} & = & v^{2}(\lambda_{2}+\lambda_{3}+\lambda_{6})\,,
\label{massashiggs}
\end{eqnarray}where $S_1$ and $S_2$ are new CP-even scalars and have masses proportional to the scale of symmetry breaking of the model $v_{\chi^\prime}$, while $H$ is the SM Higgs boson. The vev {\it v} which appears in Eq.(\ref{massashiggs}) must be equal to $246/\sqrt{2}$~GeV, in order to reproduce the right masses of the SM gauge bosons. We used in Eq.(\ref{massashiggs}) $\lambda_4=\lambda_{5}=1/4$ simply to simplify the mass terms, but we emphasize that throughout this work we performed a numerical analysis without assuming any simplifying assumption regarding the couplings. 

Besides the three CP-even scalars, a {\it CP-odd scalar} ($P_1$) remain in the spectrum with mass:
\begin{eqnarray}
M^{2}_{P_{1}} = \frac{1}{2}(v_{\chi^\prime}^{2}+\frac{v^{2}}{2}).
\label{massP1}
\end{eqnarray}

An additional complex neutral scalar also rises from the spectrum namely $\phi$, with mass given by
\begin{eqnarray}
M^2_{wimp} & = & \frac{(\lambda_{7} + \frac{1}{2} )}{2}[v^{2}+v_{\chi^\prime}^{2}].
\end{eqnarray}
 
Lastly, because of the presence of charged scalar fields in the triplet of scalars in Eq.(\ref{scalarcont}), the models contains two massive charged scalars $h_1$ and $h_2$ with masses
\begin{eqnarray}
M^{2}_{h^{-}_{1}} & = & \frac{\lambda_{8}+\frac{1}{2} }{2}(v^{2}+v_{\chi^\prime}^{2})\,, \nonumber \\
M^{2}_{h^{-}_{2}} & = & \frac{v_{\chi^\prime}^{2}}{2}+\lambda_{9}v^{2}\,.
\label{massash1h2}
\end{eqnarray}

As one can see, the scalar sector of the 331LHN model is rather rich. We have discussed and presented the mass spectrum and identified the WIMP of the model so far. Further, we will derive bounds on the dark sector by using direct dark matter detection and LHC data.


\section{Bounding the Dark Sector}
\label{darksector}
As we discussed in the previous section, the 3-3-1LHN model has a complex scalar ($\phi$) as DM. The stability of our dark matter candidate is guaranteed by a parity symmetry described in Eq.(\ref{discretesymmetryI}). In this work we revisit the DM phenomenology and derive new robust bounds on this complex DM scalar. We begin by studying the connection with Higgs.

\subsection*{Higgs Constraints}
Interestingly it has been pointed out in the literature that such complex scalar could be a potential explanation for the Galactic Center gamma-ray excess for the case that $M_{WIMP}\simeq 20$~GeV annihilating to bb \cite{331DM2}. Albeit, such a light DM particle might quite constrained in models which the DM couples directly to the Higgs. Now that the Higgs discovery has been anchored and its properties well measured at $10\%$ level, we are able to constrain in a trivial way the mass of any particle directly coupled to the Higgs by imposing the predicted branching ratio of the Higgs into new species not exceed the current bounds. Since in this
331 model the Higgs couples to all scalars in the spectrum we can trivially constraint the masses of the scalars. The masses of the scalars $P_1$, $S_1$, and $S_2$ discussed in the Section \ref{sec1} are of the order of the symmetry breaking scale of the model, $v_{\chi^{\prime}}$. As aforesaid, we are taking $v_{\chi^{\prime}}$ to be of $\sim 10 TeV$, hence the final states which contains one of these scalars are kinematically forbidden. Thus the only scalars the Higgs might decay into are the charged scalar $h_1^{\pm}$ and the dark matter candidate $\phi$ with the following decay widths,

\begin{equation}
\Gamma_{H \rightarrow 2WIMPs}= \frac{\lambda_{\phi}^2}{32 \pi} \frac{ \sqrt{M_H^2 - 4 M_{WIMP}^2}}{M_H^2},
\label{GammaWIMP}
\end{equation}

\begin{equation}
\Gamma_{H \rightarrow h_1^{+} + h_1^{-}}= \frac{\lambda_{h1}^2}{32 \pi} \frac{ \sqrt{M_H^2 - 4 M_{h_1}^2}}{M_H^2},
\label{Gammah1}
\end{equation}where,

\begin{eqnarray}
\lambda_{\phi}=\frac{-v}{\sqrt{2}(1+\frac{v^2}{v_{\chi^{\prime}}^2 })} \left( \frac{M_H^2}{v^2}+ \frac{M_{wimp}^2}{v_{\chi^{\prime}}^2} \right) \nonumber
\label{couplingfi}
\end{eqnarray}

\begin{equation}
\lambda_{h1}=\frac{-v}{\sqrt{2}(1+\frac{v^2}{V^2})} \left( \frac{M_H^2}{v^2}+ \frac{M_{h_1}^2}{v_{\chi^{\prime}}^2} \right).
\label{couplingh1}
\end{equation}

\begin{figure}[h]
\centering
\includegraphics[width=\columnwidth]{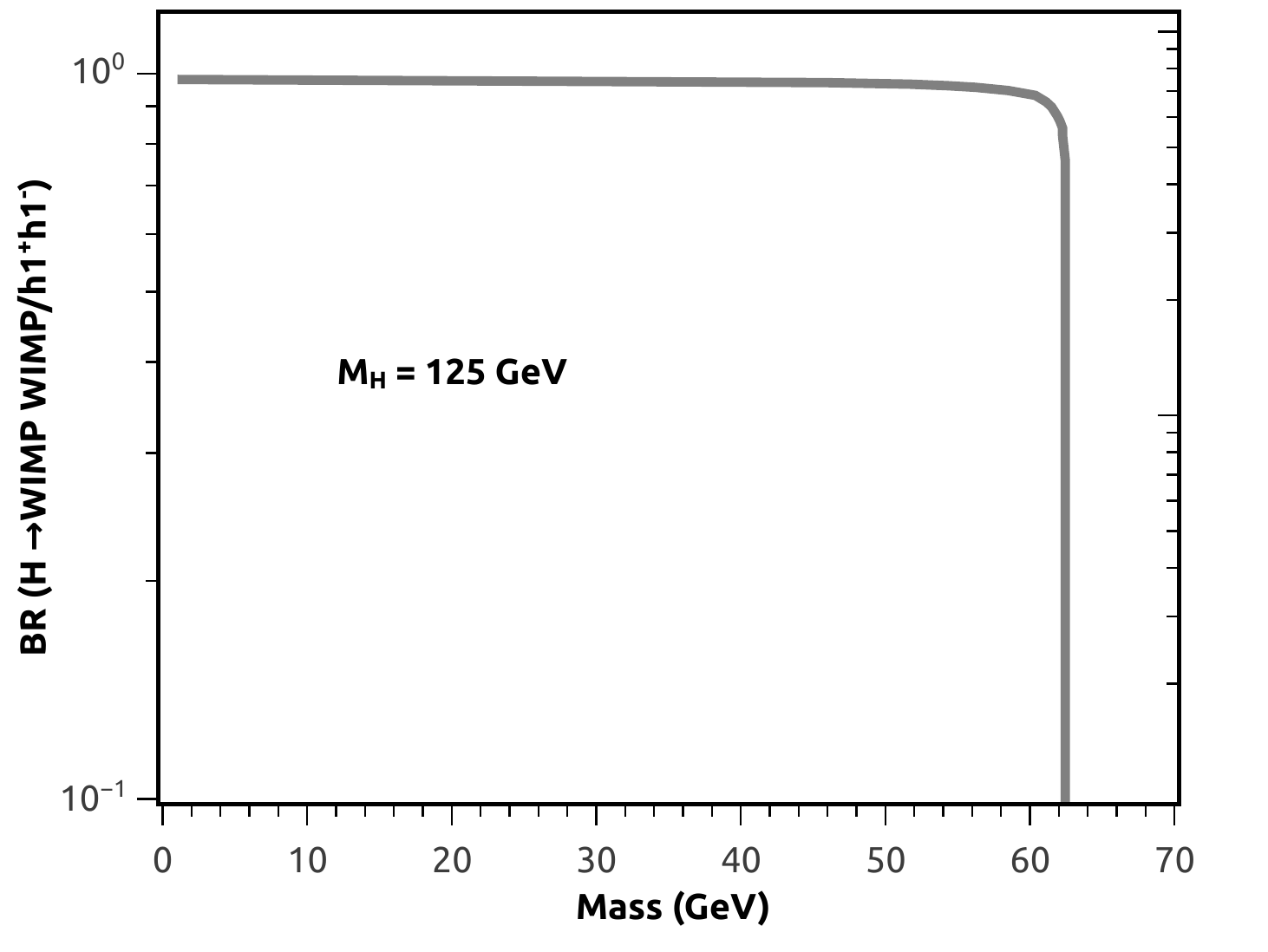}
\caption{Branching ratio of the Higgs into a pair of WIMPs/Charged scalar as a function of their masses. This result is independent of the scale of symmetry breaking of the model as long as $v_{\chi^{\prime}} \gg v$.}
\label{fig1}
\end{figure}

Notice that the Higgs width into WIMPs and charge scalar pairs are identical when the WIMP and the charged scalar masses are equal. That being said, we exhibit in FIG.\ref{fig1} the branching ratio of the Higgs in these channels as a function of their masses. Now we have derived the new Higgs decay rates some remarks are in order: 

(i) The WIMP and charged scalar decay modes overwhelm all other decay channels yielding an unacceptable branching ratio. Hence, from FIG.\ref{fig1}, we conclude that the WIMP as well as the charged higgs must be heavier than $M_H/2$, i.e $62.5$~GeV.

(ii) Differently from the so called ``Higgs portal'' \cite{hportal} where one can just use suppressed couplings to avoid the invisible width bound, in our model such alternative is not possible because the Higgs-WIMP-WIMP coupling is completely determined by the masses according to Eqs.(\ref{GammaWIMP})-(\ref{couplingfi}).

(iii) Recent results from LHC exclude branching ratios into invisible particles larger than $10\%$ \cite{higgsinvisible}, assuming the Higgs production cross section equals its SM value. In this 331 model, the new quarks do not couple to the Higgs, therefore the production cross section is the same. In other words, from precise measurements of the Higgs signal strength at the LHC we know that there is no room for a large branching ratio into missing energy in our model. Therefore, we close the light DM window in our model, namely $M_{WIMP} < M_H/2$ in order to obey the LHC bound concerning the Higgs invisible width.

At this point it is important to note that, because $\phi$ is enforced to be our DM candidate, the whole 331 mass spectrum is automatically heavier than our WIMP. Therefore this lower bound might turn out to be much stronger depending on the mass of the WIMP we are considering. Also, in order to have scalars with a mass around ~$60$~GeV some tunning is required in the coupling $\lambda_8$, according to Eq.(\ref{massash1h2}). The level of fine-tunning is dictated and proportional to the scale of symmetry breaking of the 331 gauge symmetry. Now we will turn our attention to the DM observables and derive much stronger bounds.

\subsection*{Abundance and Direct Dark Matter Detection}

\begin{figure}[!t]
\centering
\includegraphics[width=\columnwidth]{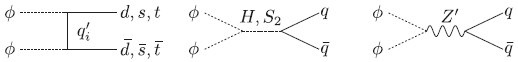}
\caption{Possible annihilation channels for a light WIMP.}
\label{figchannels}
\end{figure}

In this section we present our results concerning the abundance and direct detection observables. We have implemented the model in the Micromegas package \cite{micromegas}, and our findings are based on it. The abundance is determined by numerically solving the Boltzmann equation. Despite having many diagrams contributing to the abundance of our WIMP, we can clearly understand the role of the most relevant diagrams in Fig.\ref{figchannels}.

\begin{figure*}[!t]
\centering
\mbox{\includegraphics[width=\columnwidth]{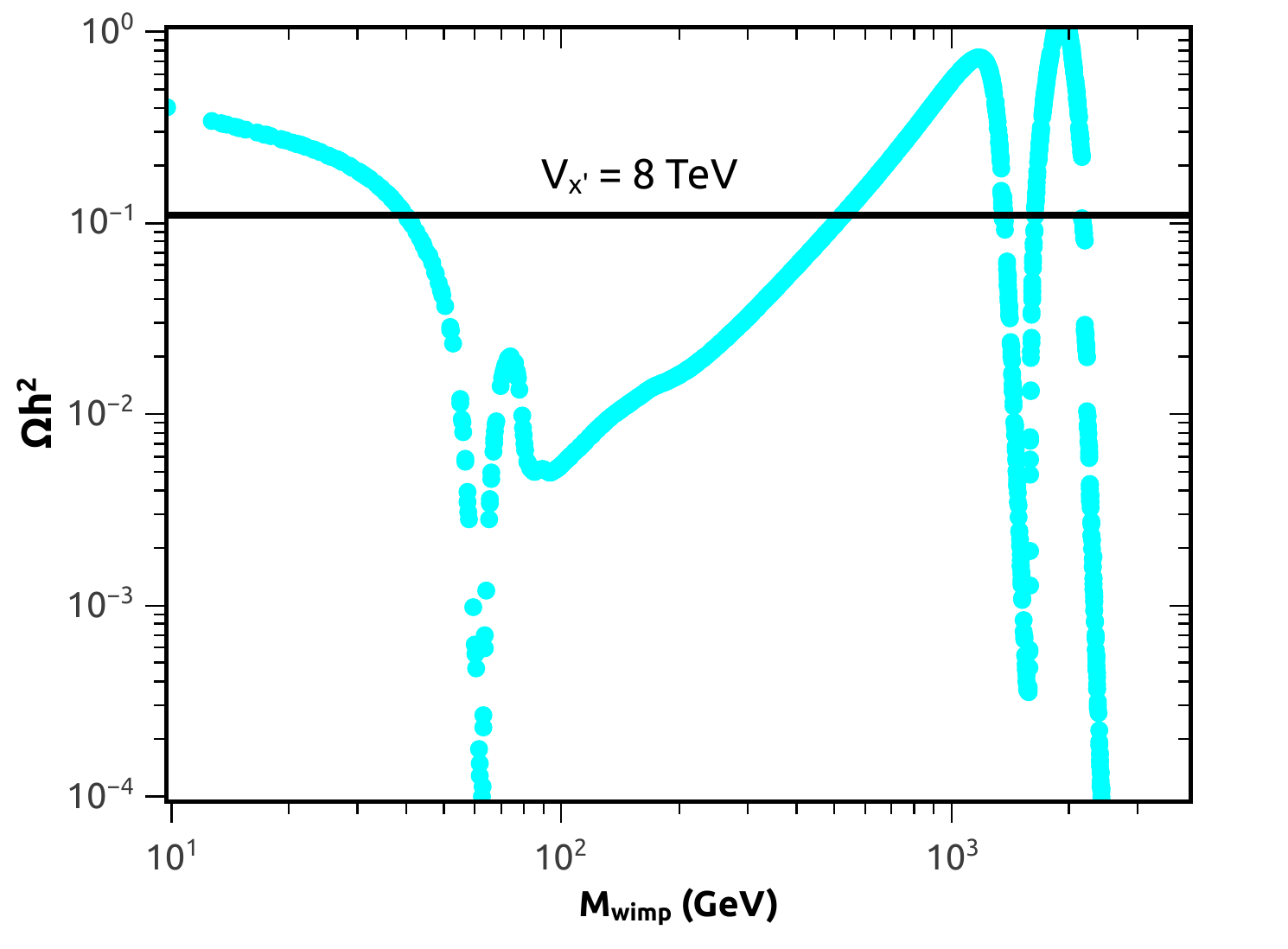}\quad\includegraphics[width=\columnwidth]{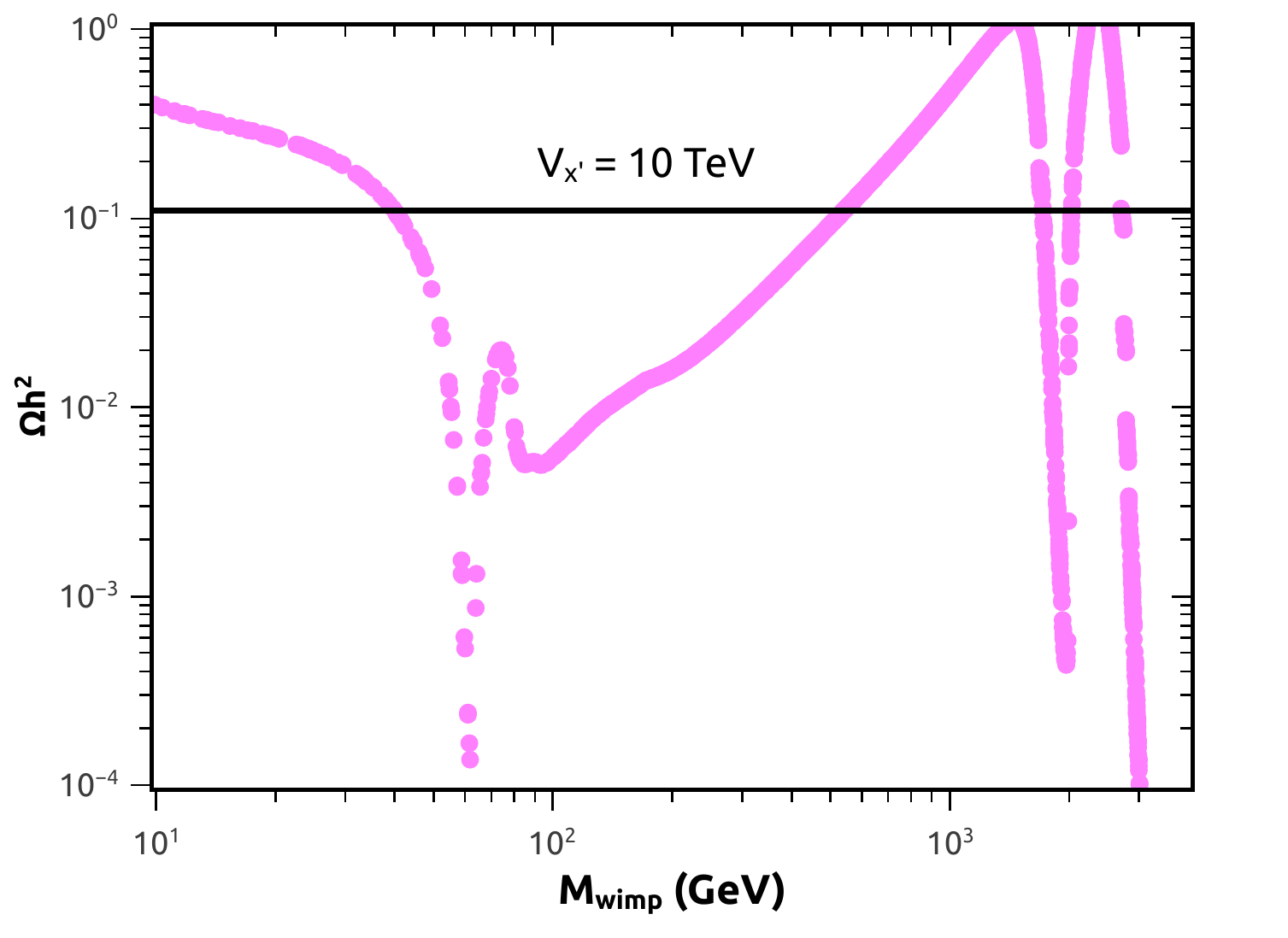}}
\caption{Abundance of our WIMP as a function of its mass for $v_{\chi^{\prime}}=8$~TeV (left panel) and $v_{\chi}=10$~TeV (right panel). Notice the resonance at $M_H/2$ happens regardless of the scales of symmetry breaking used. The second resonances occurring at $M_{wimp}=1580$~GeV and $M_{wimp}=1975$~GeV are due to the s-channel $Z^{\prime}$ mediated process shown Fig.\ref{figchannels} since $M_{Z^{\prime}}=3160,3950$~GeV for $v_{\chi}=8,10$~TeV respectively. After the $Z^{\prime}$ resonance the abundance drops again due to a resonance caused by the scalar $S_2$ whose mass is about $v_{\chi^{\prime}}/\sqrt{2}$.
}
\label{fig01}
\end{figure*}

As we know, the abundance of a generic WIMP is inversely proportional to the annihilation cross section. Hence, the resonances in the annihilation cross section set the depths of the abundance. For instance, in Fig.\ref{fig01} we have shown the abundance of our WIMP as a function of its mass for $v_{\chi^{\prime}}=8$~TeV (left panel) and $v_{\chi^{\prime}}=10$~TeV (right panel). We have drawn the horizontal line in order to easily show the parameter space that reproduces the right abundance $0.11 \leq \Omega h^2 \leq 0.12$ according to Planck \cite{planck}. One can clearly see a resonance at $M_H/2$ in both panels. This resonance remains independently of the value of the symmetry breaking used. As we increase/ decrease the latter the curve barely changes. For this reason, shifting the scale of symmetry breaking will not change our results, neither, and most importantly, the resonance at $M_H/2$. Therefore, for a light WIMP the Higgs mass determines the abundance. 

The second resonance in Fig.\ref{figchannels} occurring at $M_{wimp}=1580$~GeV  (left) and $M_{wimp}=1975$~GeV (right) is due to the s-channel $Z^{\prime}$ mediated process, since $M_{Z^{\prime}}=3160,3950$~GeV for $v_{\chi}=8,10$~TeV respectively. After the $Z^{\prime}$ resonance the abundance drops again due to a resonance caused by the scalar $S_2$ whose mass is about $v_{\chi^{\prime}}/\sqrt{2}$ according to Eq.(\ref{massashiggs}).

\begin{figure}[!h]
\centering
\includegraphics[width=\columnwidth]{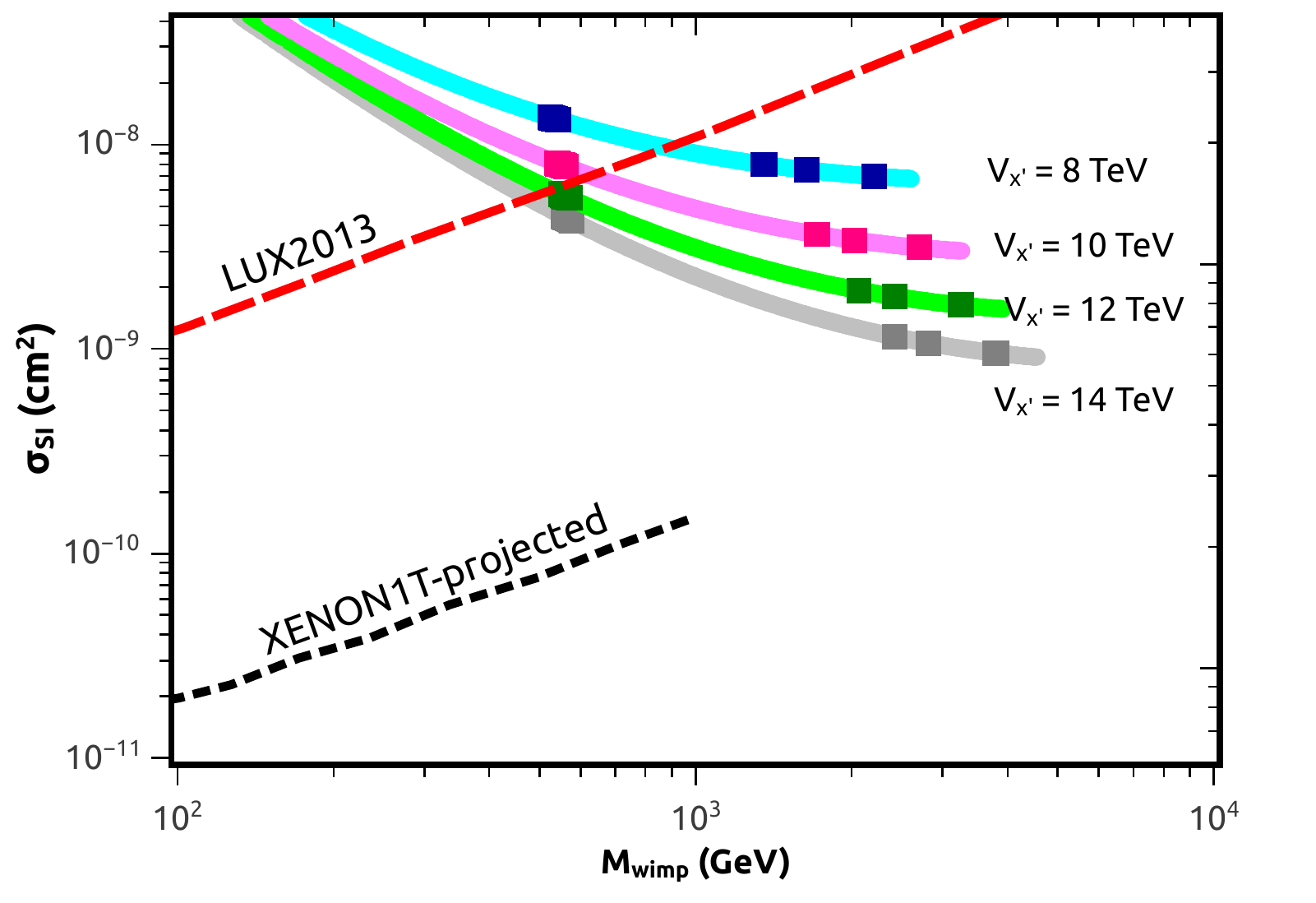}
\caption{WIMP-nucleon spin independent scattering cross section for $v_{\chi}=8,10,12,14$~TeV. The dark points delimit the parameter space that yields the right abundance in accordance with Fig.\ref{fig01}. The dashed red (black) curve is the LUX2013 (XENON1T-projected) limit. Notice that only heavy WIMPs survive LUX bounds and XENON1T projected limit will be able to exclude the dark matter mass range below $1$~TeV. }
\label{fig02}
\end{figure}

We have explained our findings regarding the DM abundance thus far, and now we will move on to the direct detection observable namely, scattering cross section. Because we have a scalar DM candidate only spin-independent scattering is induced. In Fig.\ref{fig02} we present the WIMP-nucleon spin independent scattering cross section for $v_{\chi}=8,10,12,14$~TeV. The dark points delimit the parameter space that yields the right abundance in accordance with Fig.\ref{fig01}. The dashed red (black) curve is the LUX 2013 (XENON 1 Ton projected \cite{xenon1tprojected}) bound. It means that everything above the curve is excluded by the non-observation of dark matter scatterings by the LUX (XENON1T) collaboration. 

It is obvious from Fig.\ref{fig02} that the light WIMP scenario is excluded by the current direct detection data, and for this reason our WIMP is not able to explain the few GeV gamma-ray Galactic Center excess observed in the Fermi-LAT data as claimed in Ref.\cite{331DM2}. In particular we observe that only WIMPs heavier than $1$~TeV are allowed by current data for $v_{\chi^{\prime}}=8$~TeV. Moreover, only for $v_{\chi^{\prime}}>12$~TeV WIMP masses of around $500$~GeV are not ruled out by current LUX limits. Interestingly, projected limits from XENON1T will be able to literally exclude whole dark matter mass range below $1$~TeV. 

In summary the Higgs and DM constraints, which constitute the main findings of this work are:\\
(i) Close DM masses below 500GeV;\\
(ii) Exclude light WIMPs as an explanation for the Galactic Center excess;\\
(iii) Find a lower bound of 10 TeV on the symmetry breaking scale;

Besides limits coming from the Higgs and DM abundance and direct detection observables there are relevant ones stemming from indirect detection. 

\subsection*{CMB and Indirect Detection Bounds}

The injection of secondary particles produced by DM annihilation at redshift $100 \lesssim z \lesssim 1000$ affects the process of recombination, leaving an imprint on Cosmic Microwave Background (CMB) anisotropies. Therefore, using the current measurements on the CMB power spectrum bounds on the DM annihilation cross section
have been placed namely, $\sigma v \lesssim 5 \times 10^{-27} cm^{-3}/s$ \cite{Galli:2013dna}. We have seen in the Figs.\ref{fig02} that only WIMPs heavier than $500$~GeV are not ruled out by direct detection constraints. Hence, in our model, WIMPs that yield the right abundance, i.e with a thermal cross section of $\sim 3 10^{-26}cm^3/s$ obey the CMB limits. Similarly, WIMPs that reproduce the right abundance heavier than $500$~GeV are consistent with indirect detection constraints coming Fermi-LAT \cite{fermilatbounds}.

We have seen that our model has a dark matter candidate heavier than $500$~GeV which obeys the direct, indirect dark matter detection limits as well as the collider bounds on the extra gauge bosons present in the model. Those constitute relevant findings and are the goal of this work. Further, for completeness, we comment on the charged scalar production at the LHC.

\section{Scalar production at the LHC}
\label{LHC}
 
The WIMP and charged scalars discussed previously could in principle be detected at the LHC. The detection of our WIMP at the LHC is less likely due to the featureless signal, i.e, a large amount of missing energy. Additionally, current LHC bounds on complex scalars are rather weak \cite{Willshepherd}. Hence the purpose of this section is to provide some results on the possible detection of the charged scalar at the LHC. 

We emphasize that this section serves to give a complementary information on the dark sector of the model and the proper background analysis is out of the scope of this manuscript since our main goal is the derivation of bounds coming from the Higgs and dark matter observables.

\begin{figure}[!h]
\centering
\includegraphics[width=\columnwidth]{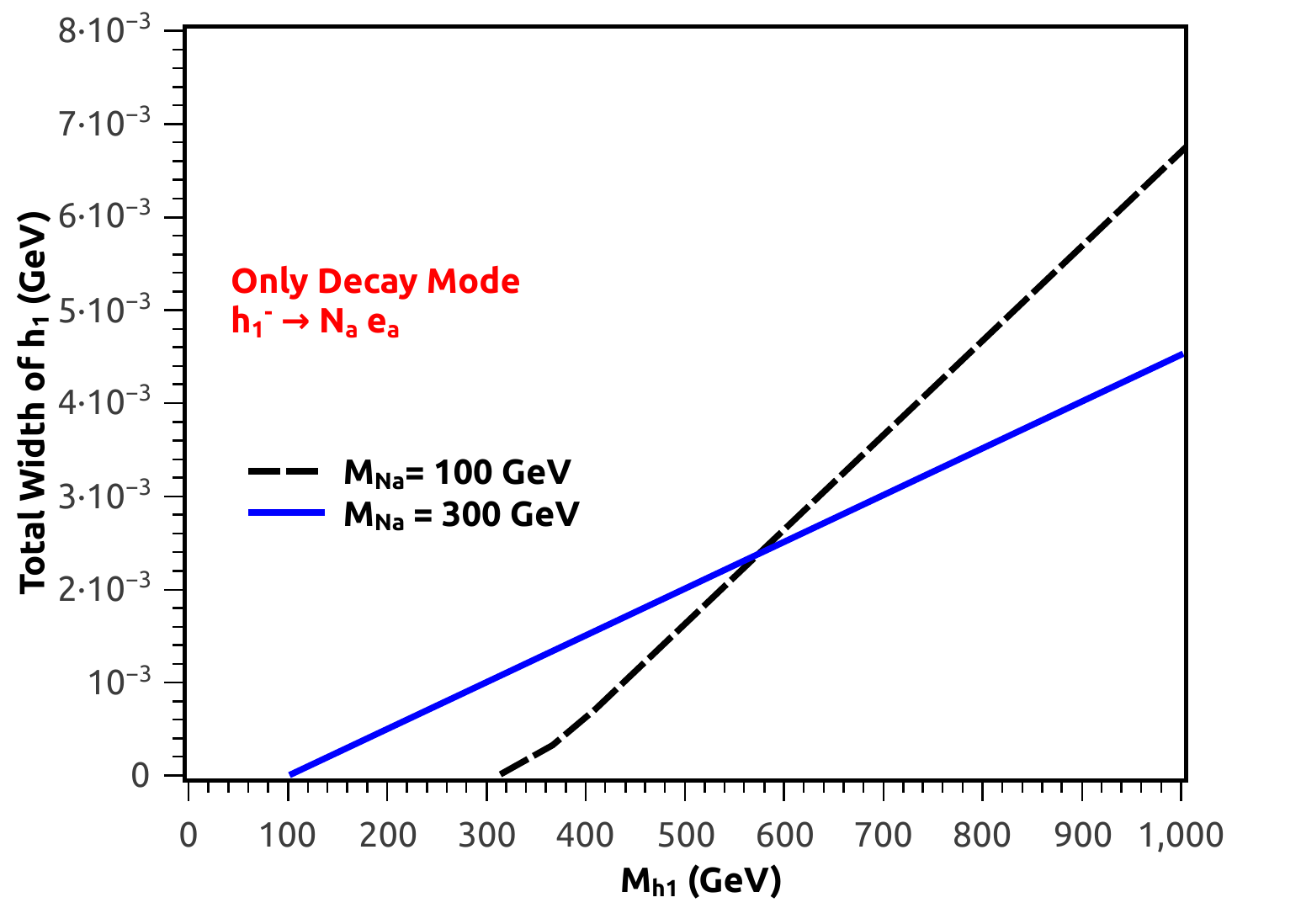}
\caption{Total decay width of the charged scalar $h_1$ as a function of its mass, for $M_{N_a}=100$~GeV (solid) and $M_{N_a}=300$~GeV (dashed). The only kinematically decay channel is $h_1 \rightarrow l_a N_a$ for $v_{\chi^{\prime}} \geq 8$~TeV. See text for details.}
\label{widthh1}
\end{figure}

\begin{figure}[!h]
\includegraphics[scale=0.35]{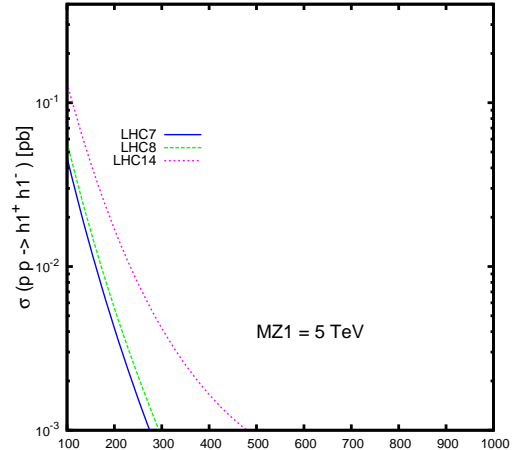}
\includegraphics[scale=0.35]{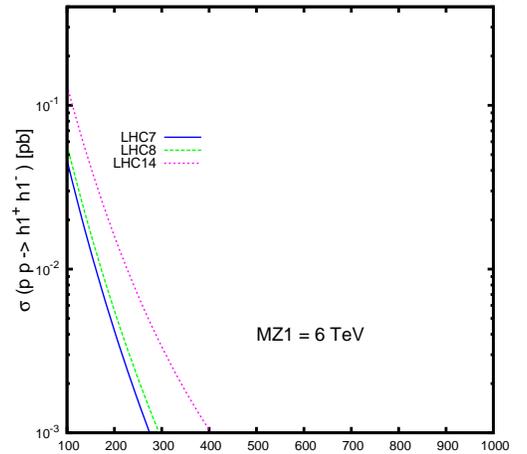}
\caption{Production cross section of the charged scalar $h_1$ at 7, 8 and 14 TeV at the LHC as a function of its mass for $M_{Z^{\prime}}=5~TeV$ (upper panel) and $M_{Z^{\prime}}=6~TeV$ (lower panel). For the regime $M_{h_1} > M_{Ni}$, $M_{Ni}$ being the masses of the heavy neutrinos the branching ratio $h_1 \rightarrow lN$ is $100\%$. From the figure we conclude that this charged scalar would have a signature similar to the W boson with higher missing energy though. Given the order of magnitude of the production cross section this charged scalar is seemingly within reach of LHC at $14$~TeV. See text for more details.}
\label{fig3}
\end{figure}

That being said, we begin showing in Fig.\ref{widthh1} the total width of these charged scalars. There we see that the charged scalars decay with a branching ratio of $100\%$ into the neutral heavy fermion (N) plus charged lepton pair (l). This feature is true as long as $M_{h_1} > M_{N_a}$, where $M_{N_a}$ are the masses of the heavy fermions which are assumed to be equal for simplicity. The coupling $h_1^+l^+N_{a}$ is proportional to the masses of the heavy fermions  and the charged leptons involved. Therefore, in the regime of degenerate heavy neutrinos masses, the $\tau N_3$ channel overwhelms the other channels.  It is important to point out tough, that for sufficient heavy charged scalars the final states $V^+Z,\; V^+Z^{\prime},\;V^+h,\;\mbox{and}\; U^0 W^+$, among others are kinematically possible. Nevertheless as we see in FIG.\ref{fig3}, once we increase the mass of the charged scalars their production cross section becomes too suppressed, making their observation quite unlikely at the LHC.

We point out that some deviations of the partonic level prediction are expected when detector effects and showering are included. Although the efficiency of the LHC detectors for events with hard electrons and muons, and large missing transverse energy can reach 96\%-99\%, the tau leptons are more difficult to detect due to the larger background from misidentified jets. Anyway, tau identification efficiency is larger than 65\% for $P_{T}^{\tau}>20$ GeV.

In Fig.\ref{widthh1} we have plotted the total width for $M_{N_a}=100$~GeV (solid) and $M_{N_a}=300$~GeV (dashed). Moreover, we have adopted $v_{\chi^{\prime}} = 10$~TeV. For such symmetry breaking scale, the remaining particles of the 331 model are heavier than $h_1$. Therefore, the charged scalars decay with a branching ratio of 100\% into the neutral heavy fermion (N) plus charged lepton pair (l). For the same reason, when $M_{h_1} < M_{N_a}$ the total decay width of the charged scalar is zero. The latter regime is problematic though, because long lived charged scalars would form the so called \emph{heavy Hidrogen} that have strong abundance limits as discussed in Ref.\cite{CHAMPS}. 

With that bear in mind, in Fig.\ref{fig3} we have computed the production cross section $\sigma (pp \rightarrow h_1^+ h_1^- \rightarrow l N_a l N_a)$ at LO, using CalhHEP 3.4.3, with CTEQ6L as the default parton distribution function, for the LHC operating with center of mass energy of $7, 8$ and $14$~TeV with $M_{Z^{\prime}}=5,6~TeV$ fixed. This production cross section is mostly driven by the $Z^{\prime}$ mass. The relevance of this particle comes from its s-channel production. Additionally, due to the decay models aforesaid the benchmark final state predicted in this scenario is the resonance production of charged leptons plus missing.

From Fig.\ref{fig3} we recognize that the charged scalar production cross section falls steeply when the charged scalar mass meets $M_{Z^{\prime}}/2$, which is what one would naively expect regarding pair production resonances. 

During the LHC Run I, from 2010 to 2012, one has reached an integrated luminosity of $L = 23\mbox{fb}^{-1}$ for center-of-mass energy $\sqrt{s}=8$ TeV in the CMS and ATLAS experiments.

According to Fig.\ref{fig3}, for the 8 TeV scenario, both production cross sections for $M_{Z_{1}}=5$~TeV and 6~TeV range up to around $\sigma = 57$ fb. Hence this new particle discovery is seemingly attainable at the LHC collider since, for the $M_{h1}=100$ GeV scenario and assuming the previous $\tau$-lepton detection efficiency, one would expect around N = 553 signal events.      

After attaining the maximum center of mass energy of 14 TeV, it is expected that the LHC will reach its design luminosity of $L = 10^{34}\;\mbox{cm}^{-2}\;\mbox{s}^{-1}$. This peak value should give a total integrated luminosity over a one year of about $40\;\mbox{fb}^{-1}$. Therefore, knowing from the Fig.\ref{fig3} that, for the 14 TeV scenario,  one would have cross sections in the order of $\sigma = 100$ fb for a charged scalar with mass of 100 GeV then we would expect to yield at least 2000 signal events just during the first year of LHC14 running. 

Moreover, in the first 10 years, the LHC shall produce a total integrated luminosity of $300\;\mbox{fb}^{-1}$, improving even more the yield of the charged scalar, mainly considering the relatively high masses scenarios. In fact, in the case of $M_{h1}=300$ GeV, with $\sigma=2$ fb, we would expect  N = 253 yielded events.

After discussing the some LHC phenomenology concerning the charged scalar production we come to our conclusions.

\begin{figure}[!b]
\includegraphics[scale=0.4]{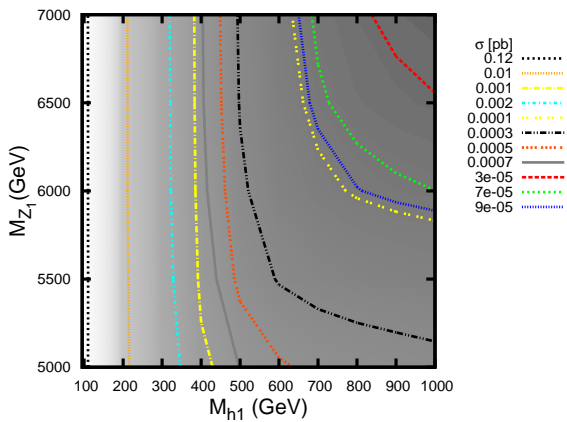}
\caption{Production cross section of the charged scalar $h_1$ at 14 TeV at the LHC as a function of the $Z^{\prime}$ and charged higgs ($h_1^+$) masses. The $Z^{\prime}$ mass drives the pair production through a s-channel diagram.}
\label{fig4}
\end{figure}

\section{CONCLUSIONS}
\label{sec:conclusion}

We have examined bounds on the dark sector of a 331 model known as 331LHN, which contains heavy neutral fermions ($N_{a}$) and a complex scalar dark matter particle in its spectrum, based on the current Higgs and direct dark matter detection data. The model is comprised of three scalar triplets, and interestingly, all of them couple to the Higgs boson. Therefore, we found a lower bound on the mass of these scalars by imposing the LHC constraints concerning the Higgs signal strength. In particular, we found that it requires the mass of the WIMP ($\phi$) and the charged scalars $(h_1^{\pm})$ to be all heavier than $M_H/2$~GeV, regardless of the coupling values used, differently from the so called Higgs portal.

We have also computed numerically the abundance and scattering cross section of the WIMP taking into account all possible amplitudes. Combining the Higgs and DM constraints we found the most stringent constraints in the literature on this model. Our main results read:

(i) Close DM masses below 500GeV;\\
(ii) Exclude light WIMPs as an explanation for the Galactic Center excess;\\
(iii) Find a lower bound of 10 TeV on the symmetry breaking scale;

Moreover, the projected XENON1T bounds are expected to fiercely rule out the entire 1GeV-1TeV dark matter region. Therefore, combining the Higgs and dark matter data, we decisively close the light dark matter window in this model and showed that the scale of symmetry breaking of this model has to live at the $\sim 10$~TeV in order to have a viable DM candidate. 

Lastly, for completeness we have computed the production cross section of the charged scalars $h_1^{\pm}$ at the LHC, which is driven by the $Z^{\prime}$ mass, and concluded that these charged scalars might be within reach of the LHC at 14~TeV as shown in FIGS.\ref{fig3} and \ref{fig4}. 

\acknowledgments

DC is partly supported by the Brazilian National Council for Scientific and Technological Development (CNPq) Grant 484157/2013-2, AXGM by the UCMEXUS-CONACyT Post-doctoral Fellowship, FSQ by Department of Energy Award SC0010107 and CNPq, and P.R.T. by CAPES. It is a pleasure to thank Alex Dias and Alexandre Alves for reading the paper and important comments. The authors also thank Patrick Draper, Will Shepherd, Carlos Pires and Paulo Rodrigues for useful discussions.

\end{document}